\documentclass[draftcls,onecolumn,11pt]{IEEEtran}

\usepackage{amsmath, amssymb}  
\usepackage{amsthm}            

\newtheorem{definition}{Definition}

\usepackage{graphicx}
\usepackage{tabularx}
\usepackage[colorlinks=true, allcolors=blue]{hyperref}
\usepackage{dsfont}
\usepackage{tikz}
\usepackage{circuitikz}
\usepackage{hvlogos}
\usetikzlibrary{calc,positioning,arrows,quantikz2,decorations.markings}
\usepackage[T1]{fontenc}
\usepackage{listings}
\usepackage{xcolor}
\usepackage{physics}
\usepackage{mathrsfs}
\usepackage{hyperref}
\usepackage{relsize}
\usepackage{mathtools}
\usepackage{algorithm}
\usepackage[noend]{algpseudocode}
\usepackage{comment}
\usepackage{soul}
\usepackage{varwidth}

\definecolor{codegreen}{rgb}{0,0.6,0}
\definecolor{codegray}{rgb}{0.5,0.5,0.5}
\definecolor{codepurple}{rgb}{0.58,0,0.82}
\definecolor{backcolour}{rgb}{0.95,0.95,0.92}

\newcolumntype{s}{>{\hsize=.25\hsize}X}
\newcolumntype{z}{>{\hsize=.45\hsize}X}

\lstdefinestyle{mystyle}{
    backgroundcolor=\color{backcolour},   
    commentstyle=\color{codegreen},
    keywordstyle=\color{magenta},
    numberstyle=\tiny\color{codegray},
    stringstyle=\color{codepurple},
    basicstyle=\ttfamily\footnotesize,
    breakatwhitespace=false,         
    breaklines=true,                 
    captionpos=b,                    
    keepspaces=true,                 
    numbers=left,                    
    numbersep=5pt,                  
    showspaces=false,                
    showstringspaces=false,
    showtabs=false,                  
    tabsize=2
}

\begin{document}

\title{Simulation of entanglement based quantum networks for performance characterizations}

\author{David Pérez Castro, Juan Fernández-Herrerín, Ana Fernández Vilas, Manuel Fernández-Veiga, and Rebeca P. Díaz Redondo \thanks{DDavid Pérez Castro, Juan Fernández-Herrerín, Ana Fernández Vilas, Manuel Fernández-Veiga, and Rebeca P. Díaz Redondo are with the atlanTTic Research Center (IC Lab), University of Vigo, Spain.
Corresponding author: Ana Fernández Vilas. avilas@uvigo.es} }

\maketitle

\begin{abstract}

Entanglement-based networks (EBNs) enable general-purpose quantum communication by combining 
entanglement and its swapping in a sequence that addresses the challenges of achieving long 
distance communication with high fidelity associated with quantum technologies. In this 
context, entanglement distribution refers to the process by which two nodes in a quantum 
network share an entangled state, serving as a fundamental resource for communication.  
In this paper, we study the performance of entanglement distribution mechanisms over a 
physical topology comprising end nodes and quantum switches, which are crucial for 
constructing large-scale links. To this end, we implemented a switch-based topology in 
NetSquid and conducted a series of simulation experiments to gain insight into practical and 
realistic quantum network engineering challenges.  These challenges include, on the one 
hand, aspects related to quantum technology, such as memory technology, gate durations, 
and noise; and, on the other hand, factors associated with the distribution process, such 
as the number of switches, distances, purification, and error correction. All these factors
significantly impact the end-to-end fidelity across a path, which supports communication 
between two quantum nodes.  We use these experiments to derive some guidelines towards the
design and configuration of future EBNs.
\end{abstract}



\section{Introduction}
In classical networks, information necessarily travels through dynamical routes from senders to receivers, which has led to the development of methods to ensure secure communication, such as encryption. However, both the transmission of information and the computational limitations of encryption are potential points of attack for eavesdroppers. These vulnerabilities have highlighted the importance of quantum communication~\cite{bennett1984proceedings, scarani2009security, kimble2008quantum}, which offers theoretical unconditional security, leveraging quantum properties such as superposition and entanglement~\cite{einstein1935can}. Unlike classical encryption, quantum communication security is not based on computational assumptions but on fundamental physical laws, like the no-cloning theorem and quantum correlations, making unauthorized interceptions easily detectable. Specifically, we consider in this paper networks based on entanglement swapping~\cite{cao2022evolution}, a quantum process that allows distant nodes to possess entangled particles without direct interaction, establishing a virtual link. These networks necessitate efficient entanglement distribution mechanisms to realize a true entanglement-based network (EBN). These mechanisms remarkably include quantum switches and quantum repeaters (or 2-switches)~\cite{Pirandola2019, briegel1998quantum}, which are crucial for constructing large-scale links and adapting to the preexisting physical environment's complex structure. Such advancements are essential for creating scalable, global, and effective quantum networks and, ultimately, the quantum internet~\cite{rohde_2021,azuma2023quantum}. However, achieving this goal remains a significant challenge, as quantum switches currently present experimental implementation restraints. This challenge has sparked increasing interest in quantum network simulators \cite{ToolsforQNet, qunetsim, SeQuEnce, 6657074qure}, such as NetSquid~\cite{coopmans2021netsquid}.

NetSquid is a discrete event simulator that operates at the physical layer of the network. By using a bottom-up approach, it can simulate the intricate quantum processes that qubits undergo, while maintaining a high degree of customization in the elements of the network and communication protocols~\cite{kozlowski2020designing}. This simulation capability is invaluable for testing network protocols and configurations, enabling the optimization of quantum networks in present times when experiments are costly and suboptimal for Noisy Intermediate-Scale Quantum (NISQ) devices. 

This work contributes to the literature by enabling the simulation of a general topology for a EBN in NetSquid, based on entanglement distribution switches, focusing on the protocols and applications that could emerge in a real network. Additionally, we have implemented a first version of E2E fidelity aware routing~\cite{zhao2022e2e}.

The rest of the paper is organized as follows. Section~\ref{sec:related} presents a review of the related literature, concluding with a discussion on how our implementation contributes to the existing body of research. Section~\ref{sec:ebn} introduces the concept of EBN, followed by Section~\ref{sec:protocols}, which provides a detailed definition of the network's components and key performance metrics are outlined. Section~\ref{sec:protocols} offers a comprehensive explanation of the simulator's elements and workflow. Section~\ref{sec:experiments} describes the experimental setup is justified and the potential applications and Section~\ref{sec:results} summarizes the obtained results. Finally, Section~\ref{sec:finale} summarizes the conclusions and outlines directions for future work.

\section{Related work}\label{sec:related}

Currently, entanglement distribution and entanglement swapping are the two key 
enabling mechanisms that are expected to be used in order to realize the 
quantum Internet. Since entanglement generation and distribution is subject to 
fundamental physical limits and has a high cost, several works have analyzed 
the achievable performance of these EBN networks. \cite{Pirandola2019} gives 
bounds on the transmission rates in a repeater-assisted quantum communications
network, while~\cite{Dai2020,Chakraborty2020} discuss entanglement distribution 
under a linear algebraic formalism, and~\cite{Fittipaldi2024} adopts a similar 
algebraic framework for the scheduling problem of entanglement resources. A 
control architecture to optimize the performance of an entanglement-generation 
switch  is presented in~\cite{gauthier2023control}, aiming to set the grounds 
for  scalable and efficient control architectures in future quantum networks.
Analytical models and results for the capacity of a single quantum switch have
been developed in~\cite{Vardoyan_QSwitch,switchcapacity,vardoyan2024bipartite} 
for swapping-based switches, and in~\cite{panigrahy2023capacity} for switches
with quantum purification. While these theoretical characterizations offer 
deep insights into the achievable rate of point-to-point quantum communications
systems, their extension to general networks seems intractable,
restricting the analysis to linear topologies~\cite{Andrade2024}, basically.

Once the entanglement resources have been created, their distribution to the 
nodes and the routing of communication requests is a problem that has been 
considered in many 
works~\cite{Caleffi2017,Li2021,Cicconetti2021,Gu2024,Sutcliffe2023}, 
with different criteria for path selection (e.g., fidelity, reliability,
fairness or maximum entanglement rate). Simultaneous concurrent requests for 
arbitrary source-destination pairs were addressed in~\cite{Shi2020,Zhang2021,Zhao2021}
for maximization of the throughput, which show that knowledge of the global
network state results in improved performance. A more realistic link model
with noise is considered in~\cite{zhao2022e2e} for end-to-end routing based on 
fidelity. However, many of these techniques are limited to specific simple
topologies or are based on heuristics, thus being suboptimal.

More recently, generalized multipartite entanglement has been proposed as a more 
effective form of generating shared states among $n > 2$ parties in a 
network~\cite{Sutcliffe2023}. Compared to bipartite entanglement and swapping,
multipartite entanglement and fusion of states probe different paths at the 
same time, and also support the deployment of multiparty quantum protocols in 
a native way. Additionally, these protocols do not have to know the exact 
route requested for end-to-end entanglement in advance, so they can attain a 
very high throughput on regular topologies~\cite{Patil2022}, independently of 
the distance. However, these newer schemes are still incapable of providing 
high entanglement rates, and are limited to servicing a single request at a time.

In view of the high sensitivity of quantum protocols to the network configuration,
i.e., the network size, the existing network paths, the amount and technology of 
the quantum memory registers at the switches, and to the physical parameters in
the links as well, simulation is necessary to help in the design of EBN networks,
since the entanglement generation rate drops quickly with the distance between
switches, the hop count, and a number of physical impairments like decoherence 
and loss. As examples, a simulation-based approach to study performance was followed 
in~\cite{vardoyan2024bipartite}, which presents a bipartite distribution network
where a real topology is analyzed by means of numerical simulation to obtain 
the capacity of the network. While its scaling is complex, this method yields 
a practical upper bound for the study of such systems. Numerical simulation 
combined with convex analysis is also the method chosen 
in~\cite{tillman2024calculating} for computing the capacity of a single switch.
For multipartite entanglement, \cite{avis2023analysis} used a hybrid
approach by combining both discrete-event simulation and Monte Carlo analysis 
to bound the performance of the central node.

In this paper, for the goal of a precise characterization of EBN networks 
under a minimal set of assumptions, we extend the preexisting modules of a 
general-purpose  discrete-event simulator of quantum networks 
(NetSquid~\cite{coopmans2021netsquid,qrepchain}). We have created a simulation
tool for networks of arbitrary topology, and using this extension along with
its related protocols, we incorporate applications that could possibly
emerge in a real setting, while supporting any kind of topology. 
Different metrics can be obtained through this tool, allowing for a variety 
of analysis to be realized. Furthermore, an E2E fidelity-aware routing 
scheme~\cite{zhao2022e2e, kar2023routing} is incorporated. Recent related works to
ours are~\cite{10621263,labay2023reducing}.  The first one also builds an algorithm 
in NetSquid to select the best possible path by estimating the fidelity 
with high confidence and very low consumption of quantum resources, 
outperforming previous methods. But, instead of using the fidelity obtained 
directly through NetSquid, it analytically simulates the fidelity as a 
Gaussian distributed variable. Labay et al.~\cite{labay2023reducing} aim at 
the resource optimization of general quantum repeater chains through NetSquid, 
utilizing state of the art parameters and protocols. Our approach allows
further degrees of freedom in defining the network configuration and the 
necessary protocols, and we demonstrate this flexibility with a study of the impact 
of technologies used in the switches and network architecture on the achievable
fidelity and on the entanglement generation rate.

\section{System Model for Entanglement Based Networks}
\label{sec:ebn}

We consider an EBN as an arbitrary topology of physical channels, networking devices (switches with multiple connections) and end nodes, where requests are generated. Requests will determine which two end nodes will communicate, and a routing algorithm will select the best available end-to-end (E2E) path using fidelity as the decision key. Bipartite states are distributed between neighboring nodes in the path, and quantum switches (QSs) will perform entanglement swapping to produce E2E entanglement.

\begin{figure}[t]
  \centering
  \includegraphics[width=0.8\columnwidth]{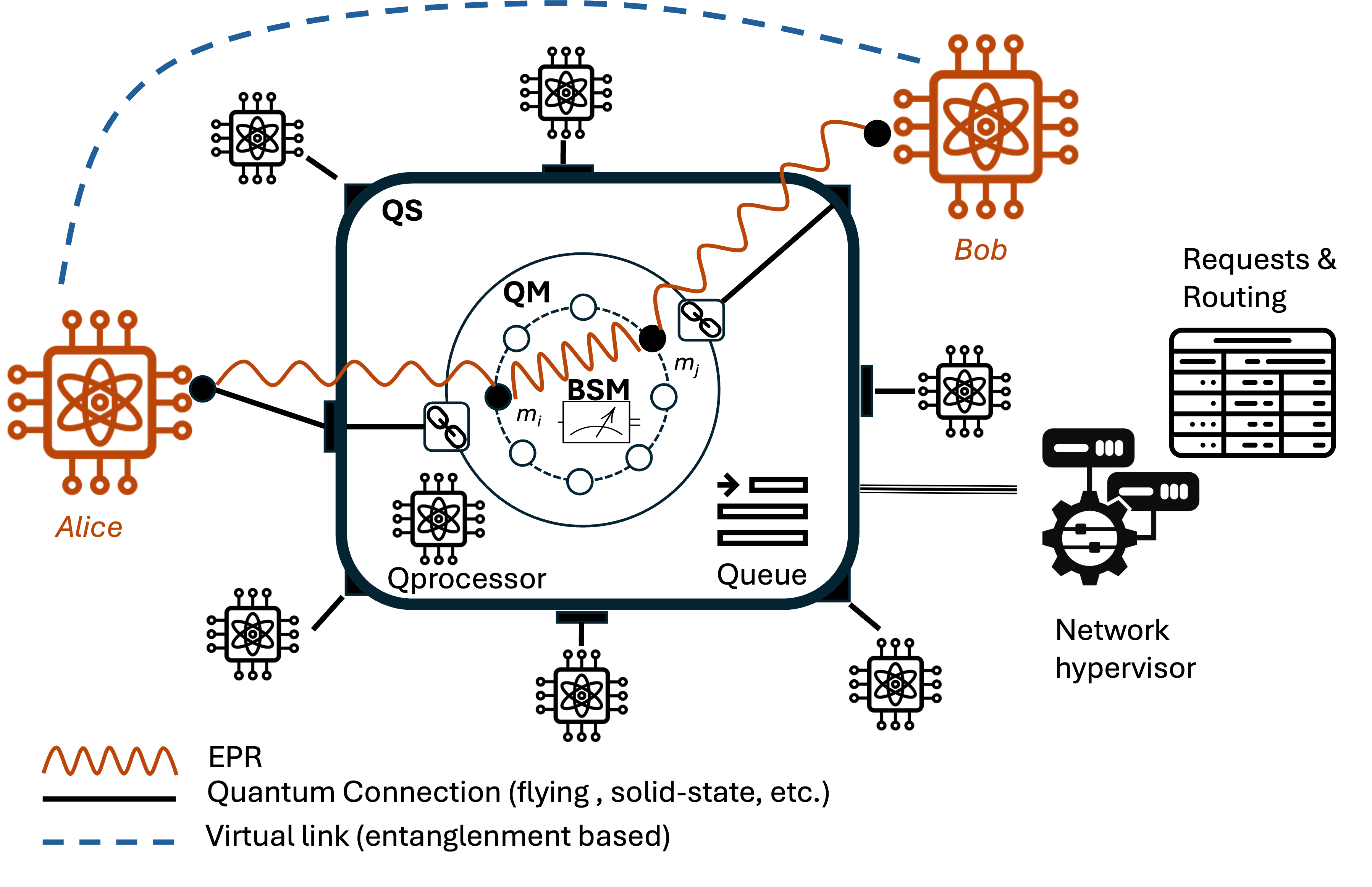}
  \caption{\label{fig:qswitch} Two nodes request a connection for  end-to-end entanglement throughout a quantum switch, which is orchestrated by the network hypervisor. The main components of a QS consist of $k$ input/output links,  a quantum processor for local operations and $m$ quantum memory units.   As  quantum processor  can only execute one operation at a time,  QSs are also equipped with a queue. Figure based on our original approach in ~\cite{PerezCastro2024}.}
\end{figure}

As in our previous work~\cite{PerezCastro2024}, we adopt  as the general model for a quantum network an abstract  weighted directed graph $G = (V, E, \{ \omega_e \}_{e \in E})$ where the  vertices $v \in V$ represent either end-nodes or intermediary nodes, the edges  $e \in E$ represent communication links (classical and  quantum) between  connected nodes, and each edge is labeled with a weight $\omega_e \geq 0$,  which in our case will be the average fidelity between the entangled pairs 
created at $e$.  Intermediary nodes are generic quantum switches   which support the creation of end-to-end entanglement. A QS is modeled (see Figure~\ref{fig:qswitch}) as a device having $k$ input/output links,  a quantum processor for local operations and and $m$ quantum memory units, where  each memory unit can store one qubit. As the quantum processor  can only execute one operation at a time, independently of the memory position involved, switch nodes are also equipped with a queue for the pending operations.

From the perspective of users,  end-nodes generate E2E Entanglement requests $\mathcal{R} = \{ r_i =
(s,d,F, \ell,\delta): i = 1, 2, \dots \} $. Given a  request $r_i$, we denote the identifiers of the source and destination nodes as $s$, $d$, respectively; $F$ is the minimum end-to-end fidelity required; $\ell$  is the maximum tolerable delay to establish the request; and
$\delta$ the desired persistence time for the duration of the entanglement. 
The special  case $\ell = \infty$ means that the request has not a setup time constraint.  
It is supposed that there is some form of end-node identification (addressing) which is 
not relevant for the contributions in this paper.

From this general topology of entanglement-based quantum networks, we are interested in studying the process of distributing entanglement all over network to satisfy the demand $\mathcal{R}$ (requests). For that we adopt a proactive routing strategy \cite{abane2024entanglementroutingquantumnetworks}, where E2E entanglement requests are received by an network hypervisor which establishes the path $p_i$ from $s$ to $d$ and orchestrate end-nodes and switches, so that the  creation of the virtual link for $p_i$ 
takes place before elementary entanglements are created. This mechanisms is  supported by the  requests and routing table depicted in Figure~\ref{fig:qswitch}.

Starting from this model assumptions, the contribution in this paper relies on the NetSquid implementation of the complete EBNs life cycle, including the physical architecture, the application requests, and the routing an virtual topology  to satisfy the demand. In
addition to the implementation of the supporting network, we have also designed and implemented a set of simulation experiments to study the figures of merit of the network. Finally, we have also simulated a set of teleportation scenarios  over an EBN. 

\subsection{Network Performance Metrics}
\label{sec:characterization}

For an EBN an a set of requests $\mathcal{R}$, the key performance metrics in the 
network overlay $G$ are capacity, fidelity, processing overhead and QBER (Quantum Bit 
Error Rate). Importantly, these metrics will be, by definition, associated to requests, 
to assess them individually and collectively and more thoroughly derive insights about
the physical network and the trade-offs between metrics. Furthermore, and for a specific
request $r_i$ in a path $p_i$,  values may vary every E2E entanglement attempt $j$. 
In the following, we use $r_i$ to denote the request to be served throughout a 
pre-assigned path $p_i$. 

First, the processing overhead $T(r_i)$ is defined as the average time interval 
elapsed from the start of the entanglement generation process until the E2E qubits 
are used for the requested application $r_i$. 
\begin{definition}
\label{def:processing-overhead}
    The processing overhead is $T(r_i) = \frac{1}{\sum_j c_j} \sum_{j} t_j(s,d)$, where, $c_j \in \{ 0, 1 \}$ accounts for every successful ($c_j = 1$) or 
    failed ($c_j = 0$) E2E entanglement generated for the request $r_i$ in the attempt $j$ 
    across the path $p_i$.  
\end{definition} 
\noindent Although the formula looks heavily simplified, $c_j$ is calculated in the end nodes, and that factor accounts for all the error sources in the network. Furthermore, $t_j$ can be broken down into individual pieces as 
\begin{equation}
    t_j = t^e_j + t^s_j +  t^{c}_j + t^p_j,
\end{equation}
accounting for entanglement generation time of the longest physical link in the path ($t^e_j$), entanglement swapping operations ($t^s_j$) and, time of propagation of the  Bell measurement at the furthest node (correction)  ($t^{c}_j$) and, if needed, purification time ($t^p_j$). 

Next, the network capacity $C(r_i)$ or throughput refers to the number of entangled states
that can be shared between the requested end nodes in $r_i$ per unit time, in this case we 
use Einstein-Podolsky-Rosen~\cite{einstein1935can} pairs (EPRs) per second. 
\begin{definition}
  \label{def:capcity} 
  The capacity of the EBN network is $C(r_i) = \frac{1}{c} \sum_j \frac{c_j}{\min{(\delta,t_j)}}$, where the denominator accounts for the fact that the processing overhead can exceed the persistence time $\delta$, and $\delta$ would take its place as the process would terminate in that case.
\end{definition}
Finally, the fidelity $F$ of each request $r_i$ is determined by the average of the fidelities of each attempt $j$. Specifically,
\begin{definition}
  \label{def:fidelity} 
  The average fidelity of a request is $F(r_i) = \frac{1}{\sum_j c_j}\sum_{j} F_j(s,d)$, where the fidelity $F_j$ of attempt $j$ is calculated as
  \begin{equation}
    F_j(s,d) = \operatorname{Tr} \biggl( {\sqrt{\sqrt{\rho_j(s,d)}\sigma\sqrt{\rho_j(s,d)}}} \biggr)^2,
  \end{equation} 
  and $\rho_j(s,d)$ is the final shared state between source and destination of the request, affected by all modeled sources of error and entanglement swapping. For the purposes of this paper, $\sigma=\ket{\Phi^+}\bra{\Phi^+}$ is the maximally entangled state, where $\ket{\Phi^+} = \frac{1}{\sqrt{2}} (\ket{00} + \ket{11})$.
\end{definition}

Fidelity is a similarity measure between quantum states that holds an exceptional value 
in quantum-information analysis and in simulations, but that is very hard to realize in
experiments, specially for distant states. To overcome this challenge, our proposed solution
for a realistic experimental setting (real or simulated) is to replace the fidelity with 
the QBER, using the latter as an indirect measure of fidelity. The procedure consists in
measuring both states at $s$ and $d$ after every E2E pair is generated, compare the
measurement outcomes ---communicating over a classical channel---, and 
counting the percentage of erroneous bits in the last node with respect to the first node. 

All the figures of merit described above try to capture the impact of physical equipment in 
the network by different error models in quantum channels (e.g. attenuation and 
depolarization) and quantum memories (e.g. decoherence and de-phasing). 

\section{Software Design and Implementation}
\label{sec:protocols}

We have used the base classes in NetSquid to implement a general topology for EBNs, 
as well as a set of simulation experiments.  A physical quantum topology consist of 
a set of quantum nodes interconnected by quantum channels where each intermediate node
(quantum switch) is equipped with a quantum processor and memory  to participate in the
corresponding swapping operations. In this framework, quantum switches act as multiple-path
repeaters by performing bipartite entanglement swapping operations on the physical qubits
stored in particular memory positions that are reserved for different paths. These switches
enable the creation of virtual links by concatenating entanglement (by swapping) from 
shorter physical links and so the creation of the weighted virtual graph $G$. 

In a potential realization od EBNs,  we envisioned a \textit{network hypervisor} (manager 
or controller)  in charge of generating and distributing entanglement for every tuple 
request $r_i = (s,d,F,\ell,\delta)$. The hypervisor has full access to the state of the
network so that memory positions in the intermediate switches can be pre-assigned to a
specific virtual path for $r_i$. Each path $p_i$  corresponds to the sequence of physical
links that, through entanglement swapping,  provides an E2E virtual link between  $s$ 
and $d$. Different requests can refer to different quantum applications, although in 
this work only teleportation is studied, $F,\ell,\delta$ are the parameters related with 
these applications. Ambitiously,  an  hypervisor should assign $p_i$ to $r_i$ guaranteeing 
and efficient management of quantum resources and ensuring reliability and scalability 
for large networks.

\begin{figure}[t]
  \centering
  \includegraphics[width=\columnwidth]{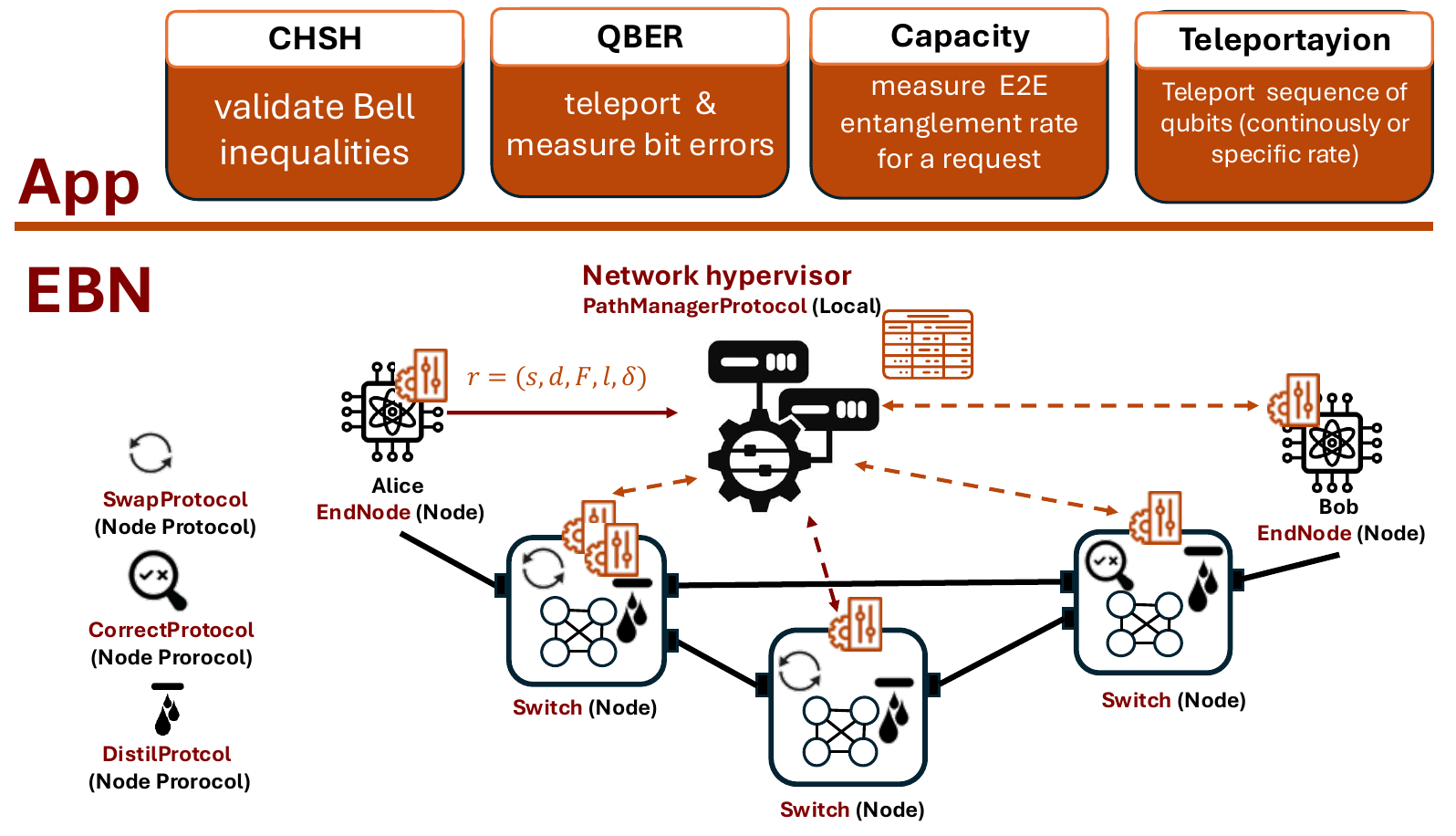}
  \caption{\label{fig:ebn-app} This figure shows the general topology for EBNs and the set of simulation experiments (applications) for this paper. The new classes (mainly nodes and protocols) are shown with the base NetSquid classes in black. A network hypervisor has also been designed and implemented to handle requests and manage E2E entanglement. Not all applications are reported in this paper.}
\end{figure}

NetSquid includes entities (classes) for the main physical elements in a quantum network, 
i.e., quantum nodes (class \texttt{Node}) and links (class \texttt{Channel}) as well as 
the corresponding error models (class \texttt{Model}). From these building blocks, network
logic is encapsulated within the \texttt{Protocol} class, which supports networking and 
final applications by fundamental functionalities such as signaling between protocols and 
life cycle management, including configuration, initialization, execution, and termination.
NetSquid network logic is deployed at two levels: \texttt{NodeProtocol}, which controls a
single network element, and \texttt{LocalProtocol}, which operates across multiple network
elements and  coordinates a set of instances of \texttt{NodeProtocol}. On top of these base classes  Figure \ref{fig:ebn-app} shows our design which is explained in the following sections \footnote{The software can be find here https://github.com/jfherral00/EBN-Simulation}

\subsection{Nodes \& Links}
We have implemented two extensions, by inheritance,  from NetSquid \texttt{Node} class:
\texttt{Switch} and \texttt{EndNode}. Regardless of the node type, they contain a set 
of quantum memory positions and a quantum processor, as described in Section~\ref{sec:ebn}.  
A \texttt{Switch} has, at least, as many memory positions as connected links, so that, 
the more memory positions the more paths involved in the virtual graph. On the other hand,
an \texttt{EndNode} contains only two memory positions to generate entangled pairs 
(mandatory if purification is  applied). 

Moreover, quantum processors at the nodes add quantum computation capabilities to perform
operations on the memory positions. As one processor cannot execute multiple operations
simultaneously, even if they affect different memory positions, a queue for pending 
operations (so requests) is needed. In  our implementation,  instruction execution times 
and dephasing/depolarization rates for each operation can be parameterized at the 
\texttt{Swicth}.

The physical topology of the quantum network is defined by the links between nodes. 
Generally, end nodes are connected to switches by two links at most, meanwhile the number 
of links between switches is only limited by memory positions. A link between a pair of 
nodes is characterized by a set of physical  parameters:  transmission delay, dephasing,
depolarization and losses. 

\subsection{Requests and sources}

To manage entanglement requests, the hypervisor handles each request $r_i$ by 
pre-assigning a $p_i$ between the between $s$ and $d$ so that EPR sources can generate
bipartite states along $p_i$. Then, entangled pairs are used for specific applications, 
such as teleportation (as in this paper), either at a defined demand rate or continuously 
if no rate is specified. The quantum state that will be teleported is stored in 
a separate dedicated quantum memory. As in NetSquid entanglement sources are located at the channel, 
we modify these sources by setting the distance from the source to the switch to zero 
to be functionally assumed as in-switch sources.
 
Finally, to accommodate fluctuations in demand, the quantum switch is equipped with
an additional memory that functions as a queue. This queue is designed to handle
situations where the entanglement demand rate exceeds network capacity, supporting 
different management strategies such as FIFO and LIFO. The queue's size is adjustable, 
and if the demand exceeds its capacity, qubits are discarded according to the selected 
strategy.

\subsection{Error Models}\label{sec:errormodels}

Noise in quantum memories and channels represents a critical bottleneck in the realization of  quantum networks. Quantum memories suffer from decoherence and dephasing while channels are affected by attenuation and depolarization, leading to significant losses during transmission. These noise sources not only reduce the fidelity  but also limit the capacity of the quantum network.  Memory positions can be affected by different decoherence models: dephasing, depolarization, or T1T2 \cite{RevModPhys.62.745} (decoherence and dephasing time), though only vacancy center memory models have been used for the results of this work. Notably, we have implemented a depolarization model, which makes use of the polarization mode dispersion time $\tau_{\text{PMD}}(L)$ of the pulse to account for our depolarization probability as a function of the channel distance, based on~\cite{Huebel2007}. For the channel depolarization, also~\cite{Huebel2007} is used; meanwhile  T1T2 models  are the usual approach for memory decoherence and depolarization, e.g., in vacancy centers, namely NV and SiV centers, each with their own attributes (gate duration, photon conversion rate, output photon wavelength, etc).  In terms of memory decoherence, state of the art technologies are compared in our work, in particular vacancy centers.

Out of these sources of error, the two primary limiting factors are fiber attenuation and memory decoherence. Both factors introduce considerable losses in quantum signals, requiring technological advancements to achieve functional large-scale quantum networks. These factors significantly impact network capacity, which is the most critical performance metric relative to classical communications, and should be prioritized in network construction, besides fidelity and time. Clearly, the mitigation of channel losses is one of the reasons for deploying quantum switches (also non-linear topologies).  The distribution of switches and the resource allocation is still an open problem, and is one of the main results in this work.  

\subsection{Simulation: Pre-assigned routes and distribution protocols}
\label{sec:simulation}

Given the infrastructure in the physical topology, the quantum dynamics in the EBN 
effectively distribute entanglement to provide E2E links. In our simulation tool,
this is done as follows. 
To start with, the virtual  graph $G$ is created from an estimated fidelity as 
weighting metric. Then, quantum protocols are installed and launched to simulate 
on Netsquid the local operations and classical communication (LOCC) as the set of 
operations and measures which support entanglement creation, entanglement swapping, 
and entanglement purification. A combination of these primitive mechanisms is thus 
arranged to simulate quantum communications.

In more detail, the EBN and the quantum communication processes are configured by mapping 
the physical topology and the associated physical parameters of the network to
the internal model, including the nodes, the channels, and the error models, as well as 
the communication requests $\mathcal{R}$. Then, the simulation of a EBN works in two 
separate stages.  First, a \emph{virtual graph} $G$ is constructed out of the physical
topology by simulating the physical links independently $n$  times to achieve a narrow
confidence interval for fidelity. These estimations of fidelity are used as the weights 
of the edges $E$ of $G$. In a second stage,  \emph{application stage}, the  network 
hypervisor is initiated to monitor the network state and to consequently establish  a 
proper scheduling of the requests  $r_i$ according to the path $(s,d)$,  the fidelity 
threshold $F$, and the maximum  waiting time $\ell$.  The hypervisor, with full 
knowledge of the network, calculates the optimal routes and the number of purification
rounds, and finally orchestrates all protocols to execute the requests)concurrently. A
sequence of requests of the same type forms a \emph{quantum application} between the
two end-nodes, e.g., qubit teleportation. More examples will be explained below in 
Section~\ref{sec:experiments}.

For managing the requests in our EBN, we apply a routing protocol similar to the
end-to-end) fidelity aware routing in~\cite{zhao2022e2e}, implemented using
the estimated average fidelity as link weights in the virtual graph $G$. 
Dijkstra's minimum-cost algorithm is used to select the path with  maximum E2E fidelity,
for each  $r_i$ before executing the quantum applications and the network protocols.
This optimal path is simply the one that can fulfill the minimum E2E fidelity $F$ for 
a request $r_i$. If there are multiple available paths with fidelity exceeding $F$, 
the path with the highest fidelity is always preferred. Once the paths have been a
already selected for all requests $\mathcal{R}$, quantum applications are concurrently
executed, contending for the entanglement resources. In the case that a target fidelity
$F$ cannot be achieved, entanglement purification is activated continuously: we use 
DEJMPS distillation~\cite{deutsch1996quantum}) up until one of these two conditions is 
met (i) $F$ is attained, so $r_i$ is processed, or (ii) the maximum waiting time $\ell$  
is exceeded, so $r_i$ is aborted and the execution of this request fails due to 
insufficient resources. 

This behavior is deployed by the network hypervisor throughout a set of quantum protocols 
as follows. Given the set of requests $\mathcal{R}$, the  hypervisor independently 
instantiates a \texttt{PathManager} protocol for each request $r_i$. During graph
construction, \texttt{PathManager} gathers network information to establish the path for
$r_i$, whereas during  the application stage, it waits for an external protocol to
ultimately ask for an entangled pair between $s$ and $d$. Then, \texttt{PathManager} 
coordinates three sub-protocols (swapping, purification and correction) to produce an 
entangled pair between $s$ and $d$ with enough fidelity $F$ across the assigned path.
The sub-protocols work as explained next.
\begin{itemize}
\item \texttt{SwapProtocol}. There is an instance of this protocol in every switch in 
the path for $r_i$. If purification is required, \texttt{PathManager} launches two 
instances of \texttt{SwapProtocol} instead, that is, one more to support entanglement
purification. In general, when qubits are detected in the assigned  memory positions, 
switches perform a Bell measurement  and send the results  to the destination  via the
classical channel. 

\item \texttt{DistilProtocol}. When entanglement swapping is complete, 
\texttt{PathManager} sends a signal to execute as many purification rounds as are 
associated with the $r_i$ according to the ore-assigned path. In our experiments
\texttt{DistilProtocol} implements DEJMPS purification. 

\item \texttt{CorrectProtocol}. This protocol runs only on the destination node. When 
all the corrections have been received (as many as there are switches in the path), 
it applies them to the qubit in memory to create the entangled pair. If purification 
is needed, there will be two of these protocols associated with the demand.   
\end{itemize}
Upon completion, invoking request (application) is notified about the availability of
an entanglement with the characteristics in $r_i$ so that  the application is ready to run.
Different applications can have different particular metrics associated, not just referring 
to the aforementioned key metrics in Section~\ref{sec:ebn}, such as the teleported states 
per second, the QBER for QKD applications, etc.

 \section{Experimental Design: Network Performance \& teleportation}
\label{sec:experiments}

In this Section, we present the built-in experiments designed and implemented on 
top of extended entities and protocols. In a first group, we aim to study  how the 
quantum error models affect the network when multiple error sources are considered, 
and we focus on the two limiting factors of a quantum network: memory decoherence and 
channel attenuation. This involves selecting the most suitable quantum technology 
for memories and analyzing the impact of the number of switches between end nodes 
on network performance. In a second group of experiments, our goal is to compare
the efficiency of different network applications, contrasting standard teleportation 
with error correction codes.

\subsection{Performance modeling of quantum switches}
\label{sec:performance:experiments}

A first set of experiments in this group focuses on studying the impact of 
different physical technologies used in a quantum switch on fidelity $F(r_i)$,
capacity $C(r_i)$ and processing time $T(r_i)$. To elaborate a useful engineering
model of a quantum switch, we propose to analyze numerically these variables:
\begin{enumerate}
\item Fidelity vs. Storage Time, $F(\mathrm{T1})$. We assess the impact of quantum memory 
storage time  (T1) on fidelity. To this end, three network configurations are simulated, 
and the fidelity of stored states over time is measured. This allows us to determine
the storage time required to maintain high fidelity and to understand the effect of swapping 
operations on fidelity degradation.

\item  Capacity vs. Gate Duration, $C(G)$. This experiment examines the capacity of 
a single quantum switch under different traffic loads by varying the gate duration. 
With this, we can understand the network's capacity to handle multiple requests 
depends on the duration of the quantum gates, and identify the threshold at which 
capacity saturates.

\item  Processing overhead vs. Photon Loss, $T(E)$. This experiment examines how 
the probability of photon loss, caused by various error sources, affects processing time. 
By comparing technologies such as SiV and NV centers, we analyze how different error 
rates influence the time required to establish entanglement, providing insights into 
the robustness of various quantum memory technologies.

\item Capacity/Fidelity vs Distance, $C(d)$. This experiment examines several topologies 
for a specific source-to-destination distance and evaluates their effects on capacity and
fidelity. It highlights the relative impact of decoherence and depolarization, emphasizing 
the critical factors that shape overall network performance.
\end{enumerate}
Finally, we have implemented two additional simulation settings to study the impact of 
memory technologies and number of intermediate nodes  in the  overall network performance.
\begin{enumerate}
\item  Capacity vs Memory Technologies (SiV and NV), $C_m$. By measuring the capacity 
of networks utilizing SiV and NV memories over varying distances, we identified which
technology achieves higher entanglement rates and is better suited for practical 
applications. This experiment also investigates the impact of adding switches on capacity, 
isolating the primary mechanisms contributing to photon loss.

\item  Capacity/Fidelity vs Hops, $C(h)$. We compared the impact of varying the number 
of switches between end nodes on the fidelity of entangled pairs and overall network 
capacity. This evaluation included analyzing the trade-off between minimizing the number 
of switches to reduce operational complexity and increasing them to combat decoherence 
over longer distances, thereby maximizing fidelity.
\end{enumerate}

\subsection{Quantum Teleportation}
 \label{sec:teleportation:experiments}
 
The most fundamental application for evaluating the efficiency of a quantum network 
involves counting the entangled pairs received to determine the network capacity. 
In this process, a pair is requested from the network layer, and upon the successful 
preparation of entanglement, its fidelity is measured against the Bell states 
$\ket{\Phi^+}$ or $|\ket{\Psi^+}$,  after which the state is discarded. Additionally, 
qubits at the end nodes can be measured to calculate the QBER. These metrics, combined 
with the entanglement generation time, offer a comprehensive characterization of the 
network's performance for a given topology and physical layer.

After an entangled pair is shared between the end nodes, the qubit stored at the end nodes
can be used to perform an application. One application of particular interest, implemented 
in this study, is state teleportation. In this process, an entangled pair is requested, 
and the qubits specified in the request are teleported using the shared entangled pairs. 
Three variants of this protocol are implemented: (i) \textit{Teleportation}, this variant 
does not specify a demand rate, meaning the teleportation rate matches the transmission 
rate. Qubits are created and teleported as soon as entangled links become available. 
(ii) \textit{Teleportation with demand}, this variant allows for a fixed uniform demand 
rate to be specified in the request; if the demand exceeds the network’s capacity, 
a transmission buffer is used to manage the overload. (iii) \textit{Logical Teleportation}, 
this variant incorporates error correction using Shor's $[9, 1]$ single-qubit error 
correction code~\cite{shor1995scheme} (other error correction codes could be programmed).

The network metrics generated include the number of teleported qubits, the mean 
transmission time, and the fidelity of the teleported states. For the 
\textit{Teleportation with demand} variant, additional metrics such as buffer queue 
size and the number of discarded qubits at the end of the simulation are reported. 
To analyze this application scenario, the following alternatives have been prepared 
and executed.
\begin{itemize}
\item Error correction vs. purification (EC/P). We compare the effectiveness of error 
correction and purification techniques by assessing their impact on fidelity, network 
capacity, and processing overhead. This analysis allowed us to identify which approach
delivers superior performance across various network configurations and operational 
conditions.
    
\item Capacity/Fidelity/Overhead vs Hops  (CFO/H). We conducted simulations using 
fixed network configurations to evaluate key performance metrics, including fidelity, 
capacity and processing overhead. By systematically varying the number of switches and
inter-node distances, we identified the most efficient configurations tailored to 
specific network scenarios. This experiment provides a clear  and targeted assessment 
of network performance under controlled conditions.
\end{itemize}

\section{Results}
\label{sec:results}

This Section presents the results from the simulation experiments, summarized in 
Table~\ref{table:experiments}. Performance results for quantum switches and networks are 
presented and discussed in Section~\ref{sec:results:network:performance}, with 
particular attention given to the trade-off between channel distance and the number of 
switches between end nodes. For memory decoherence, we compare various state-of-the-art 
quantum memory technologies to identify the most suitable option for large-scale networks 
in Section~\ref{sec:results:network:memorytechnologies}. Fidelity and processing overhead
results are reported in Section \ref{sec:results:network:intermediatenodes}, as these 
are also important factors in network performance. Finally, based on the insights gained 
from the previous results, we will simulate different teleportation scenarios 
(Section~\ref{sec:results:network:teleportation}).

\begin{center}
    \begin{table}[t]
    \caption{Outline of the experiments and its purpose.\label{table:experiments}}
    {\small
    \begin{tabular}{llp{0.56\columnwidth}} 
    & Focus & Description \\ \hline
     \multicolumn{3}{c}{{\bf Performance Experiments}}\\ \hline
    \multicolumn{1}{l}{$F(\mathrm{T1})$} & Memory Noise & See the different regimes that are present for a network configuration, given T1 cannot be arbitrarily large.\\ \hline
    \multicolumn{1}{l}{$C(G)$} & Gate Duration & Understand under which circumstances gate duration affects the capacity of the network. \\ \hline
    \multicolumn{1}{l}{$T(E)$} & Channel Noise 
    & Observe the linearity (or lack of) of the duration of the process with the probability of losing a photon entering the fiber, to establish its relevance. \\ \hline
    \multicolumn{1}{l}{$C(d)$} & Distance & See the effect of the distance and the number of intermediate nodes. \\ \hline
    \multicolumn{1}{l}{$C_m$} & Memory Technology & Decide the best quantum memory technology with respect of the figures of merit \\ \hline
    \multicolumn{1}{l}{$C(h)$} & Number of switches  & Analyze how capacity is affected by the physical infrastructure with the adopted error models. \\ \hline
    \multicolumn{3}{c}{{\bf Teleportation}}\\ \hline

    \multicolumn{1}{l}{EC/P} & Fidelity Preservation & Compare the different methods to mitigate fidelity losses. \\ \hline
    \multicolumn{1}{l}{CFO/H} & Number of switches  & Analyze how the number of intermediate nodes affects all figures of merit in a fixed network. \\ \hline
    \end{tabular}
        }
    \end{table}

\end{center}

\begin{figure}[t]
    \centering
    \includegraphics[width=\linewidth]{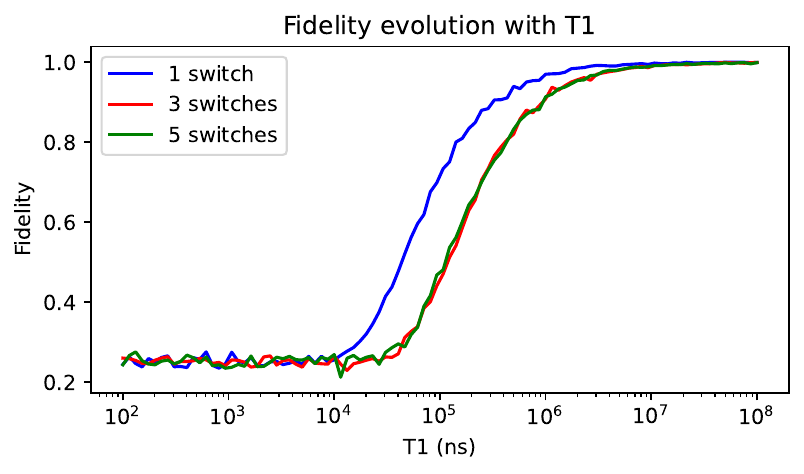}
    \caption{(Color online) Fidelity of shared states between end nodes for 3 different configurations, with fixed E2E distance of 10 km. Capacity and processing overhead can be considered constant. X axis in logarithmic scale.}
    \label{fig:fidt1}
\end{figure}

\subsection{ Network Planning: Performance of quantum switches and EBNs}
\label{sec:results:network:performance}

One of the main features of the simulator is its usefulness to run different 
parameterized simulations to assess the protocol metrics under the changes of a 
certain component. For instance, we can see in Fig.~\ref{fig:fidt1} how  fidelity 
varies in three different networks as a function of the storage time (T1) of the memories. 
For storage times under the swapping protocol time, basically all states are lost and we 
find the expected fidelity for a random Bell state, $1/4$. This swapping time also explains
the spacing between the three networks, as it is proportional to the amount of swapping
operations performed. There is a transition phase when T1 is of the same scale as the 
swapping operations time, and when T1 is orders of magnitude greater the fidelity stabilizes
at $1$. 

\begin{figure}[t]
    \centering
    \includegraphics[width=\linewidth]{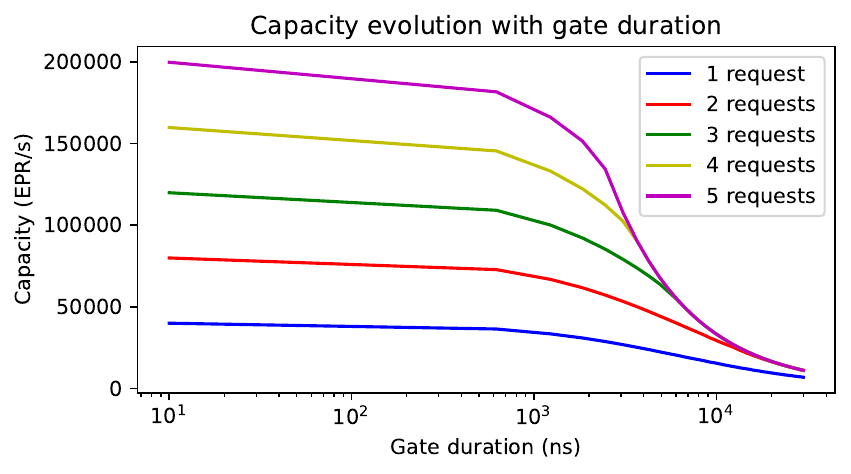}
    \caption{(Color online) Network capacity for 5 different configurations, with fixed E2E distance of 5 km. Fidelity is kept constant at $\sim 1$ and time scales linearly with gate duration. The quantum memories utilize NV centers. Memory decoherence and channel attenuation omitted. X axis in logarithmic scale.}
    \label{fig:capacityvsgates}
\end{figure}

We can also analyze capacity performance by varying the gate duration of the memories, 
as shown in Fig.~\ref{fig:capacityvsgates}. In this case, capacity of a single switch 
with different traffic loads is studied. The observed effect is influenced solely by 
the duration of the gates, as decoherence and attenuation are set to zero. As expected, 
in any configuration, as the operation time of the gates increases, the capacity decreases
because the average time for swapping operations increases. For short durations, the switch 
is very efficient, exhibiting additive behavior where capacity scales with the number of
concurrent requests. However, as the duration of the instructions increases, concurrency
increases, and capacity converges to a constant value. When the gate duration increases, 
the maximum capacity that must be delivered becomes constant and must be distributed among 
the different requests. In the configuration shown in Fig.~\ref{fig:capacityvsgates}, up to 
a gate duration of approximately 1,000 ns, the switch remains very efficient, and the 
capacity increases as requests increase. Beyond 1,000 ns, the total capacity decreases 
until it converges to a constant value when the instructions require more than 10,000 ns 
for execution.

\begin{figure}[t]
    \centering
    \includegraphics[width=\linewidth]{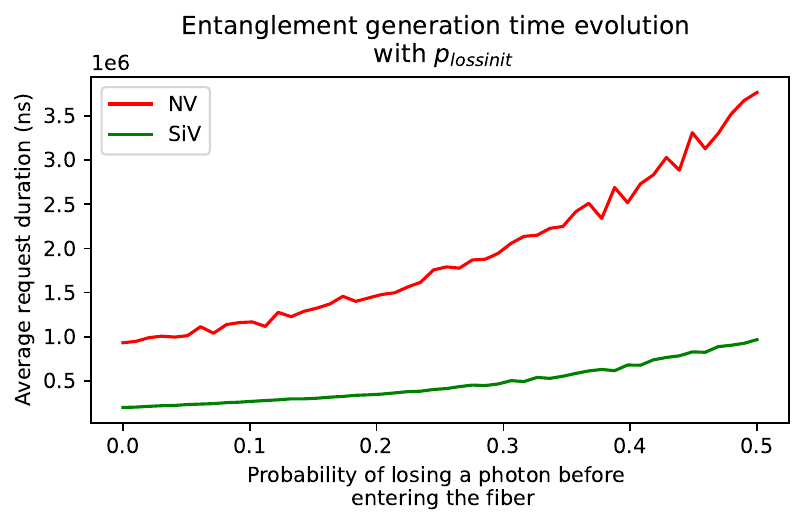}
    \caption{(Color online) Request duration for 2 different configurations, with fixed E2E distance. X axis in logarithmic scale.}
    \label{fig:timevlosses}
\end{figure}

Another interesting examination is the time dependence of a variable, in this case, 
the probability of losing a photon when it enters the fiber, or any other source of 
error besides fiber attenuation. This dependence is shown in Fig.~\ref{fig:timevlosses}.
Naturally, with a higher probability of losing a qubit, the average time to establish
entanglement increases. These results are notable due to the different scaling presented
by the two studied technologies, SiV and NV centers.

\begin{figure}[t]
    \centering
    \includegraphics[width=\linewidth]{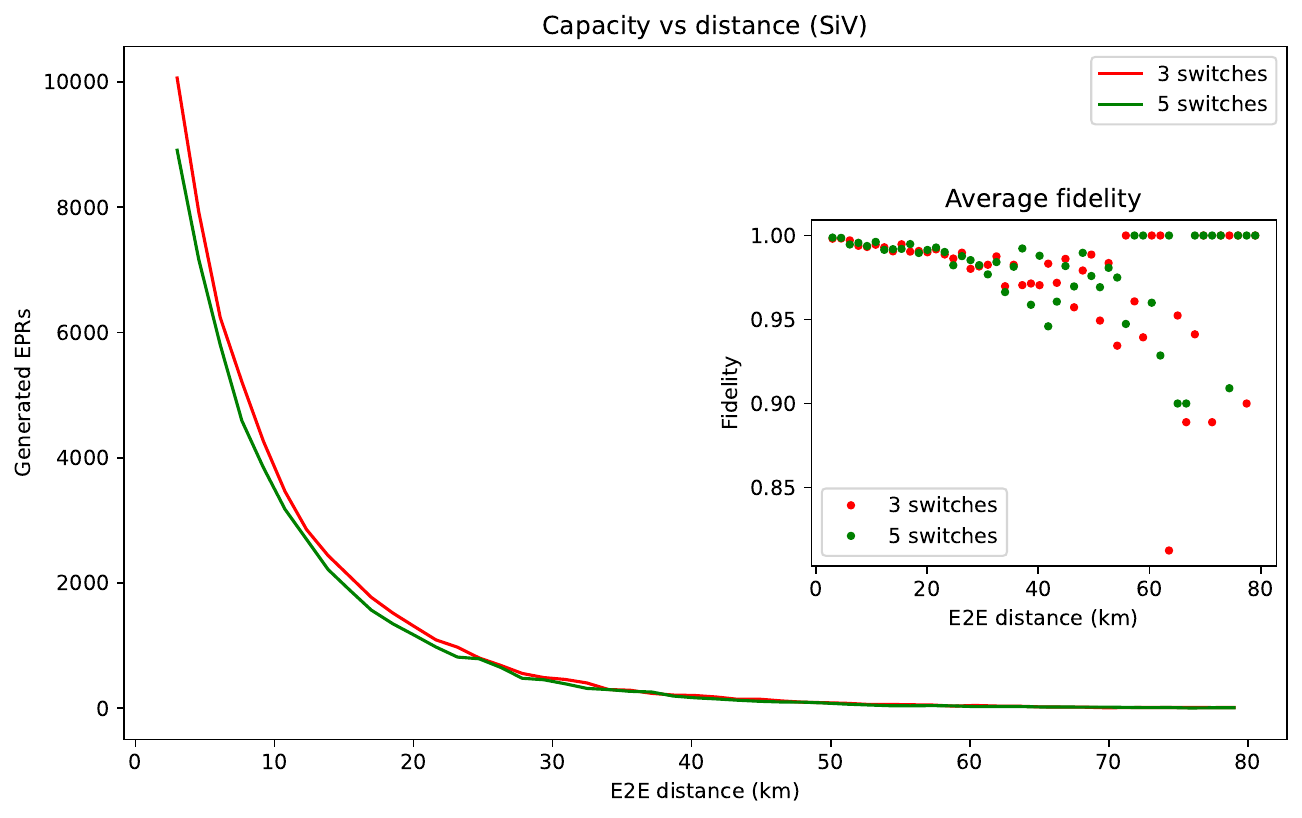}
    \caption{(Color online) Main plot: Capacity of the network as a function of the total network distance for two different network configurations. \\Inset: Average fidelity from the requests managed in the main plot.}
    \label{fig:capandfidinset}
\end{figure}

Furthermore, visualizing two of these metrics simultaneously can provide a clearer
interpretation of the obtained results. For instance, Fig.~\ref{fig:capandfidinset} 
plots capacity and fidelity against the total network distance for two different
configurations, with three and five switches respectively. This figure highlights the
intrinsic difference in the scales of decoherence and depolarization effects, showing 
that the decrease in capacity is orders of magnitude larger than the decrease in fidelity.
This disparity highlights the more significant impact of decoherence on network performance.

However, specific values and optimization strategies will be discussed in detail in the 
following sections, where we will focus on optimizing quantum memories and channels. By
addressing these key components, we aim to enhance overall network performance and better
understand the trade-offs involved in different configurations. This three-metric approach 
not only helps in identifying critical performance bottlenecks but also guides the 
development of more robust quantum network solutions.

\subsection{Technologies in quantum memories}
\label{sec:results:network:memorytechnologies}

\begin{figure}
    \centering
    \includegraphics[width=\linewidth]{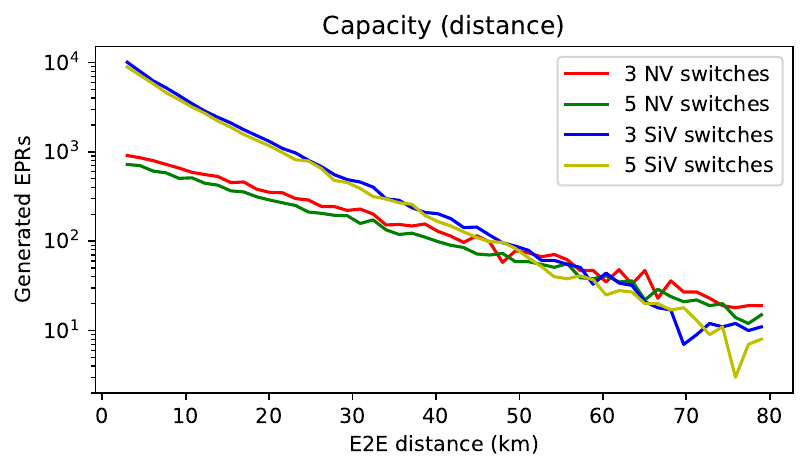}
    \caption{(Color online) Capacity as a function of the total network distance for 4 different network configurations. Y axis in logarithmic scale.}
    \label{fig:NVvsSiVcapacity}
\end{figure}

Our first study involves comparing two selected technologies for what the most 
plausible model for quantum processors is today, vacancy centers: SiV and NV.
Fig.~\ref{fig:NVvsSiVcapacity} shows the capacity results for these two technologies,
indicating that SiV memories are significantly more effective at achieving high 
entanglement rates for distances up to 40 km. This is attributed to the shorter one-qubit 
gate duration in SiV memories (as low as 1 ps, see e.g.~\cite{Becker2016}). The slight
decrease in capacity that occurs when inserting additional switches for a fixed E2E 
distance corroborates that the main mechanisms of photon loss are due to the coupling 
to the fiber, and not attenuation and memory decoherence. Distances greater than 
approximately 20 km do not provide sufficient entanglement rates for practical applications,
and will be regarded as the minimum effective distance for capacity from now on.

\subsection{Number of intermediate nodes}
\label{sec:results:network:intermediatenodes}

In addition to memory optimization, we can compare the effect of inserting multiple 
switches for a fixed E2E distance vs. minimizing the switches in between end nodes and 
how they change the fidelity of the entangled pairs, to optimize the behavior of the 
network in addition to taking into account the effect of distance to the generation rate. 
We can see in the inset of Fig.~\ref{fig:capandfidinset} that there is no statistical 
evidence to opt for either configuration below 30 km range, as basically there are no 
high rates of depolarization before decoherence already consumes all qubits. Having this 
limit overlap with the one mentioned in the capacity section allows the system to avoid
(mostly) undergoing purification processes, which increase the processing overhead by 
orders of magnitude. Lastly, time is also individually studied. Intuitively, adding a 
switch increases processing overhead, because not acting on the flying qubit is less 
time consuming than an entanglement swapping operation. However, capacity is 
dramatically decreased without switches.

\begin{table}[t]
    \caption{Capacity, processing overhead and mean fidelity for a network of fixed E2E distance of $60$ km. In this case the most efficient solution would be to minimize the switches. A performance dip at 3 switches can be seen, explained by the losses and depolarization peak due to maximum processing overhead. Rows marked with \textbf{*} exceed the minimum effective channel distance.    \label{tab:60km}}
    \begin{center}
    \begin{tabular}{cccc} \hline
         No. of switches & 
         Capacity & Processing & Avg. fidelity \\
         & (EPRs/s) & overhead (ms) \\ \hline
         5 & $43$ & $23.12$ & $1.00$  \\ \hline
         4 & $41$ & $24.01$ & $1.00$ \\ \hline
         3 & $35$ & $27.96$ & $0.94$ \\ \hline
         2 & $41$ & $23.15$ & $1.00$ \\ \hline
         1$^*$ & $50$ & $19.43$ & $1.00$ \\\hline
    \end{tabular}
            \end{center}
\end{table}

\subsection{Achievability of Quantum State Teleportation on Networks}
\label{sec:results:network:teleportation}

After conducting these tests and adopting the aforementioned optimal mechanisms, 
the quantum network infrastructure is considered optimized in terms of capacity, 
end-to-end (E2E) fidelity, and processing overhead. This optimization allows for 
real applications involving quantum information transmission and teleportation to be 
executed effectively.

\begin{figure*}[t]
    \centering
    \includegraphics[width=\linewidth]{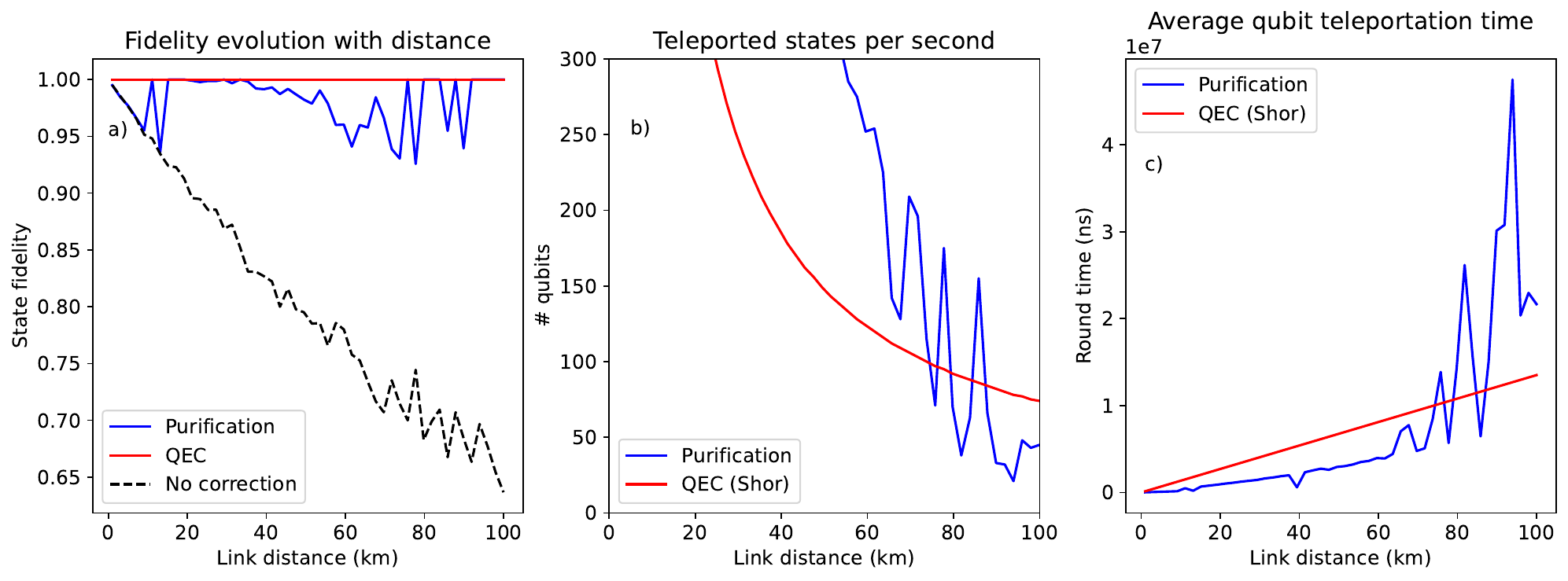}
    \caption{(Color online) Three different results extracted from different teleportation applications. \textbf{a)} Evolution of the fidelity when increasing the link distance. \textbf{b)} Evolution of the capacity when increasing the link distance. \textbf{c)} Average processing overhead as a function of the link distance.}
    \label{fig:teleportation_comp}
\end{figure*}
\begin{figure}[t]
    \centering
    \includegraphics[width=0.9\linewidth]{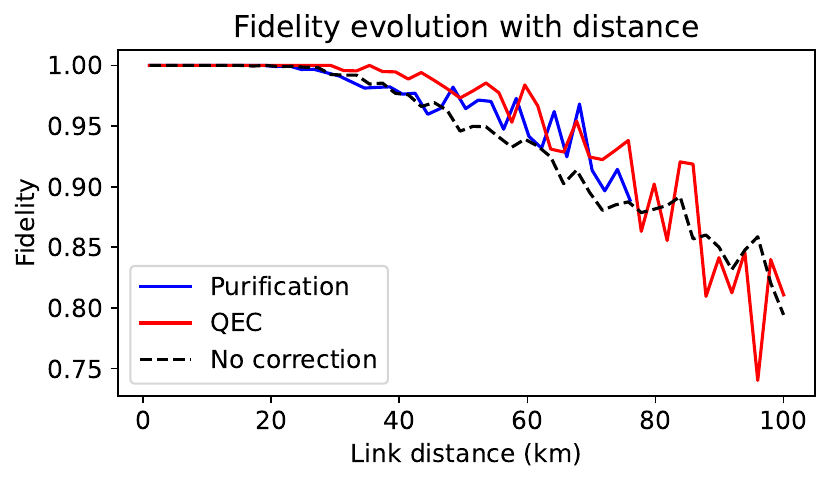}
    \caption{(Color online) Evolution of the fidelity when increasing the link distance for different teleportation applications.}
    \label{fig:fidelity_correction}
\end{figure}

For actual applications, two instances are presented. First, using the evolution method 
for the simulator, the mechanisms of error correction and purification are compared. In 
this analysis, purification of end states is activated for fidelities lower than 0.95 and 
is directly compared with Shor's code error correction. This comparison is illustrated in
Fig.~\ref{fig:teleportation_comp}, where the three different approaches ---purification,
Shor's correction, and no correction--- are evaluated in the presence of a dephasing 
error model in the $Z$ basis ($R = 500$ Hz), based on three metrics. In subfigure (a), a
clear improvement in fidelity is observed when implementing quantum error correction (QEC) 
and purification, as opposed to uncorrected states. The uncorrected states do not achieve 
the minimum required fidelity for channel lengths over 10 km and are excluded from further
analysis. In sub-figures (b) and (c), for distances previously considered practical,
purification results in higher capacity and less time per teleported qubit. However, 
this advantage diminishes rapidly at around 70 km, which may have implications for networks
different from those considered in this work. A similar study can be done for general 
faulty channels, as in Fig.~\ref{fig:fidelity_correction}. When purification is activated, 
the state degradation in the memories is very high due to the time consumption of the
purification process, and fidelity loss is not compensated. In this case, there is not a 
clear method that improves the rest in terms of fidelity, and so the best method is 
not correcting the qubits, as it is the less resource-intensive.

\begin{table}[t]
    \caption{Capacity, processing overhead and mean fidelity for a network of fixed E2E distance of $30$ km. In this case the most efficient solution would be to minimize the switches. A performance dip at 4 switches can be seen, explained by the losses and depolarization peak due to maximum processing overhead.    \label{tab:30km}}
\begin{center}

    \begin{tabular}{cccc} \hline
         Number of switches& Capacity & Processing & Avg. fidelity \\
         & (EPRs/s) & overhead (ms) & \\ \hline
         5 & $578$ & $1.73$ & $0.98$ \\ \hline
         4 & $499$ & $2.00$ & $0.98$ \\ \hline
         3 & $664$ & $1.50$ & $1.00$ \\ \hline
         2 & $727$ & $1.37$ & $0.99$ \\ \hline
         1 & $775$ & $1.29$ & $1.00$ \\ \hline
    \end{tabular}      
\end{center}
\end{table}

On the other hand, one can also use the fixed parameter approach to obtain fast results 
for a fixed configuration of the network and requests. In Table~\ref{tab:60km}, the 
results for a fixed simulation are presented in 3 different network configurations, only
varying the amount of switches and the inter-node channel distance, with a fixed E2E 
distance of 60 km, to see how the sources of error interact with each other. In this case, 
the less efficient configuration up to 5 switches is found using only 3 switches, and the 
best involves 1. In this case efficiency applies to capacity and processing overhead, as
fidelity presents no significant change. Additionally, the same approach has been presented
for a network of 30 km, in Table~\ref{tab:30km}. In this case, the least optimal
configuration is for 4 switches, manifesting the intricate nature between all the 
physical processes involved and the heavy network configuration dependence of the results.

\section{Conclusions and future work}
\label{sec:finale}

This study has provided a comprehensive analysis and optimization of quantum networks,
focusing on physical components such as quantum memories and channels, as well as their 
impact on network performance. Through detailed simulations and characterizations, 
several critical insights and conclusions have been drawn.

First, the characterization of the network and its components revealed how different 
parameters affect key figures of merit. We found that memory storage time (T1) 
significantly influences fidelity, with shorter storage times leading to higher state 
loss and lower fidelity. As storage times increase beyond the duration of swapping 
operations, fidelity stabilizes. Additionally, the capacity performance was shown to 
be dependent on gate duration, where shorter gate durations enhanced the capacity 
efficiency of the network switches, but increased durations led to a decrease in capacity 
due to the longer operation times.

Later, our comparative analysis of SiV and NV vacancy center technologies indicated that 
SiV memories are more effective in achieving high entanglement rates up to 40 km. This 
is attributed to the shorter one-qubit gate duration in SiV memories. The study 
highlighted that coupling to the fiber, rather than attenuation and memory decoherence, 
is the primary mechanism of photon loss. Therefore, SiV memories were identified as the 
more suitable technology for large-scale quantum networks, particularly for distances 
up to 20 km, which was established as the minimum effective distance for maintaining 
practical entanglement rates. Furthermore, optimizing the channel configurations 
involved examining the trade-offs between the number of switches and channel distance. 
It was found that there is no significant statistical preference for either configuration
below a 30 km range, as depolarization effects do not significantly impact fidelity 
before decoherence consumes the qubits. The overlap of this limit with the capacity 
section allows the system to avoid purification processes, thereby reducing processing
overhead. The study concluded that adding switches increases processing overhead but is
necessary to maintain capacity, especially over longer distances.

Finally, for an optimized network, which incorporates the best practices from our memory 
and channel studies, was tested for real applications involving traffic and teleportation. 
The comparison of error correction mechanisms, such as purification and Shor code,
demonstrated that purification resulted in higher capacity and less time per teleported 
qubit for distances up to 70 km in channels with only Z-error. Beyond this range, the
advantage diminished, indicating the need for different strategies for longer distances.

Overall, this paper showcases the complex interplay of various factors in quantum 
network performance and demonstrates how this simulation tool elucidates these 
intricacies. By characterizing network components, the optimization of quantum
components is feasible for a specific set of network parameters, and evaluating 
practical applications, we have identified critical performance bottlenecks and 
provided guidance for developing more robust and efficient quantum networks. For future
work, the addition of other memory technologies could improve the performance of 
the network, such as superconductors. Another point is the costly use of purification
mechanisms, which could be overcome by the implementation of other techniques such as
optimistic entanglement pumping.

\section*{Acknowledgments}
This work has been supported by project QURSA, under grants TED2021-130369B-C31, TED2021-130369BC32, TED2021-130369B-C33 funded by MCIN/ AEI/ 10.13039/ 501100011033 and by the “European Union NextGenerationEU/ PRTR”. The work is also funded by the Plan Complementario de Comunicaciones Cu\'anticas, Spanish Ministry of Science and Innovation (MICINN), Plan de Recuperación NextGenerationEU de la Unión Europea (PRTR-C17.I1, CITIC Ref. 305.2022), and Regional Government of Galicia (Agencia Gallega de Innovación, GAIN, CITIC Ref. 306.2022). Additionally, it also has been funded by the Galician Regional Government under project ED431B 2024/41 (GPC).

\bibliographystyle{ieeetran}

\end{document}